\documentstyle[aps]{revtex}
\begin{document}
\input psfig.sty
\draft

\title{\bf The Role of the D$_{13}$(1520) Resonance in $\eta$
Electroproduction}

\author{R. M. Davidson, Nilmani Mathur and Nimai C. Mukhopadhyay}

\address
{Department of Physics, Applied Physics, and Astronomy \\
Rensselaer Polytechnic Institute \\
Troy, New York 12180-3590}


\maketitle
\vskip 12pt

\begin{abstract}
We investigate the electroproduction of $\eta$ mesons below a center
of momentum energy of 1.6 GeV, with particular emphasis on the roles of the
$N^*$(1535) and $N^*$(1520) resonances. Using the effective Lagrangian
approach, we show that the transverse helicity amplitude of the
$N^*$(1535) can be extracted with good accuracy from the new eta
electroproduction data, under reasonable assumptions for the strength
of the longitudinal helicity amplitude. In addition, although the
differential cross section is found to to have a small sensitivity
to the $N^*$(1520) resonance, it is shown that a recently completed
double polarization experiment is very sensitive to this resonance.
\end{abstract}

\vspace{12pt}

  
Electroexcitation of nucleon resonances
($N^{*}$ states) is a clean way of studying
the structures of 
nucleons and their excited states. Novel experimental facilities with 
polarized electron beams, available at
accelerators like CEBAF at the Jefferson Lab (J-Lab),
and corresponding developments in the polarized target technology, along
with outstanding possibilities of 
the large solid-angle detectors (like the one 
in the Hall B of J-Lab), make the prospects of extracting electromagnetic 
amplitudes for the $N\rightarrow N^{*}$ excitations as functions of 
four-momentum transfer, Q$^{2}$, a realistic one. Indeed, the first results of
such experiments are already coming on line \cite{arm,fro}.

We have explored elsewhere prospects of $\eta$ photoproduction
as a way of studying both the $N^*$(1535) (or S$_{11}$)
resonance \cite{bud} and
the difficult to access resonance $N^*(1520)$ \cite{nil} (or D$_{13}$).
Moving away from the real photon point,
most of our current phenomenological knowledge about the Q$^2$
dependence of the S$_{11}$ and D$_{13}$ helicity amplitudes comes from the
analyses of Burkert \cite{burk} and Stoler \cite{rep}. For the S$_{11}$,
Q$^3 A_{1/2}^s$ starts to scale at about 5 GeV$^2$, while for the
D$_{13}$, the ratio
\begin{eqnarray}
{|A_{1/2}^d |^2 - |A_{3/2}^d |^2 \over
|A_{1/2}^d |^2 + |A_{3/2}^d |^2 } \nonumber
\end{eqnarray}
starts to approach unity around 3 GeV$^2$, both results in agreement with the
pQCD counting rules \cite{carl}. However, the D$_{13}$ seems to be
disappearing much faster than the S$_{11}$ with Q$^2$.
Thus, it is not clear that the
D$_{13}$ is, in fact, behaving according to pQCD\footnote{It should be
noted that for Q$^2$ $\widetilde{<}$ 3 GeV$^2$, the
behavior of $|A_{1/2}^d |^2 /|A_{3/2}^d |^2$ can
also be reproduced in constituent quark models \protect\cite{qu}.}
expectations, that is, it
is not certain that Q$^3 A_{1/2}^d$ is scaling at these Q$^2$ values.
Indeed, the D$_{13}$ has never been studied at high Q$^2$ ($\widetilde{>}$
3 GeV$^2$).
As the Q$^2$ value at which the pQCD predictions become reliable is
hotly debated in the literature \cite{pqcd}, the exceptions to
scaling should play a significant role in resolving this issue. As we will
show, eta electroproduction can provide important constraints for the
N-D$_{13}$ electromagnetic transition amplitudes.

Our work should also have an impact on distinguishing among the
variety of baryon models that have been discussed in the 
literature, starting from the old non-relativistic model \cite{quark},
SU(6) analyses \cite{su6},
relativized models~\cite{capstick}, quark models involving meson and
gluon exchanges between quarks \cite{riska}, quark models with strong
emphasis on relativity \cite{rel}, models emphasizing large N$_c$ \cite{nc},
and so on. All these approaches are 
attempts to model QCD in the context of hadron spectroscopy. Our work
here should motivate these approaches to compare with the sensitive 
observables coming from the new electromagnetic studies in the baryon
sector.

The purpose of this Letter is to investigate the role of the D$_{13}$
in various eta electroproduction observables.
As the older
data \cite{bonn}, taken at Bonn, DESY, and NINA, are not of good quality,
the original effective Lagrangian analysis \cite{ela} did not include
the D$_{13}$ resonance. In addition, no D$_{13}$ contribution was
considered in the analysis \cite{arm} of the recent J-Lab data.
Given that recent {\it photo}production
data \cite{photod} are sensitive to the D$_{13}$, it is important to examine
its role in {\it electro}production.
Thus, we shall use our effective Lagrangian approach to analyze the new
J-Lab data for eta electroproduction at Q$^2$ values of 2.4 and 3.6
GeV$^2$.  We will
determine the constraints provided by these \begin{it}differential
cross section\end{it} data on the D$_{13}$ helicity amplitudes, $A_{1/2}^d$,
$A_{3/2}^d$ and $S_{1/2}^d$, and suggest new experiments that show
greater sensitivity to these amplitudes than the current data. In particular,
based on our fits to the
differential cross section, we predict that a recently
completed double polarization experiment at J-Lab \cite{stoler} is
extremely sensitive to the D$_{13}$ helicity amplitudes. Finally, our work
will provide a test of the model dependence
in extracting the dominant helicity amplitude, $A_{1/2}$ of the
S$_{11}$ ($A_{1/2}^s$) from the data.

The recent J-Lab experiment \cite{arm}
measured the differential cross-section for $\eta$ electroproduction 
in the W = $\sqrt{s}$ range from 1490 MeV to 1590 MeV,
and at Q$^{2}$ values of
2.4 and 3.6 GeV$^2$. To a large extent, the data are independent
of $\theta$ and $\phi$, suggesting the dominance of the S$_{11}$
resonance in this reaction. However, there are minor deviations
from angular uniformity hinting at reaction mechanisms other than the
S$_{11}$ contribution. Taking guidance from what has been learned
at the real photon point \cite{nil},
we have analyzed
these data using the effective Lagrangian approach, which consists of, 
in the tree approximation,
the $s$- and $u$- channel nucleon, S$_{11}$ and D$_{13}$ exchanges,
and the $t$-channel vector meson ($\rho$ and $\omega$) exchanges.
The nucleon, S$_{11}$ and vector meson exchanges have been discussed
in Ref.~\cite{bud} and need not be discussed in detail here. The
pseudoscalar $\eta$NN coupling constant, $g_{\eta NN}$,
is taken to be 1.88, and the
relevant combinations of vector meson couplings are taken to be
\begin{eqnarray}
\lambda_{\rho}g_v^{\rho} + \lambda_{\omega}g_v^{\omega} &=& 5.9 \; ,
\nonumber \\
\lambda_{\rho}g_t^{\rho} + \lambda_{\omega}g_t^{\omega} &=& 17.5 \; ,
\end{eqnarray}
where $\lambda_i$ is the $V\gamma \pi$ coupling constant and
$g_v$ and $g_t$ are the vector and tensor coupling constants to the nucleon,
respectively.

For the S$_{11}$ contribution, we take the $\eta$N, $\pi$N and $\pi\pi$N
branching ratios to be 0.5, 0.4 and 0.1, respectively, and the total
width to be 150 MeV, which are close to the preferred values found
in Ref.~\cite{arm}. The PDG \cite{pdg} estimates the total width to
be in the range from 100 to 250 MeV, and the $\eta$N branching ratio to be
between 0.35 to 0.55. In this work, we are primarily interested in the
D$_{13}$ contribution and defer discussion of the effect of these
uncertainties on the determination of $A_{1/2}^s$ to a later
publication. We note, however, that the quantity
\begin{equation}\label{psi}
{ A_{1/2}^s \sqrt{\Gamma_{\eta}} \over \Gamma_T }
\end{equation}
can be determined nearly model-independently \cite{bud} from the data.
Thus, to a good approximation, the effect of choosing a different width
or $\eta$N branching ratio on $A_{1/2}^s$ can be determined from (\ref{psi})
by an appropriate scaling. Indeed, for the results of the three fits
given in Table I of Ref.~\cite{arm}, $A_{1/2}^s \sqrt{\Gamma_{\eta}}$
is nearly constant.

For the D$_{13}$ exchange, the strong and electromagnetic effective
Lagrangians are \cite{ela,dmw,dav}:
\begin{eqnarray}
L_{\eta N R} &=& {g_{R}\over \mu} \bar{R}^{\mu}
\theta_{\mu\nu}(Z) \gamma_{5} N \partial^{\nu} \eta \, + \, {\rm h.c.}
\; , \\
L^{1}_{\gamma N R} &=& {i e G_1 \over 2 M} \bar{R}^{\mu}
\theta_{\mu\nu}(Y) \gamma_{\lambda}
N F^{\lambda\nu}\, + \, {\rm h.c.} \; , \\
L^{2}_{\gamma N R}  &=& - { e G_2 \over 4 M^{2}} \bar{R}^{\mu}
\theta_{\mu\nu}(X) 
 (\partial_{\lambda}N) F^{\nu\lambda} \, + \, {\rm h.c.} \; , \\
L^{3}_{\gamma N R}  &=& - { e G_3 \over 4 M^{2}} \bar{R}^{\mu}
\theta_{\mu\nu}(V)
N (\partial_{\lambda} F^{\nu\lambda}) \, + \, {\rm h.c.} \; ,
\end{eqnarray}
where $M$ is the nucleon mass, $\mu$ is the eta mass, and
the tensor $\theta_{\mu\nu}(C)$ is defined as follows\footnote{We take
A = -1 in the spin-3/2 propagator \protect\cite{bdm}.}
\cite{bdm}:
\begin{equation}
\theta_{\mu\nu}(C) =  g_{\mu\nu} - [{1 \over 2} (1 + 2C)]
\gamma_{\mu} \gamma_{\nu} \; .
\end{equation}
The gauge couplings $G_i$, linearly related to the D$_{13}$
helicity amplitudes, and
the off-shell parameters \cite{bdm}, $V$, $X$, $Y$, and $Z$, are {\it a priori}
unknown and are determined from the fits to the data.
The $\eta ND_{13}$ coupling constant, $g_R$, can be determined once
the total width and $\eta N$ branching ratio of the D$_{13}$ are specified.
We take the total width to be 125 MeV and the $\eta N$ branching ratio
to be 0.1\%, that is, the D$_{13}$ $\eta N$ partial decay width is 0.125 MeV.
According to the PDG \cite{pdg}, the D$_{13}$ width is quite well determined,
being in the range from 110 to 135 MeV. On the other hand, the $\eta$N
branching ratio is quite uncertain. It is known to be small and
nonzero \cite{nil}, but at this point no reasonable error can be given
to it. The $\pi$N and $\pi\pi$N branching ratios are taken to be 0.6
and 0.4, respectively.

Since the D$_{13}$ is a $J^P$=3/2$^-$ state, the multipolarities of
the $\gamma ND_{13}$ transition can be electric dipole (E1),
magnetic quadrupole (M2) and, if the photon is virtual, Coulomb
dipole (C1). The relevant resonant electroproduction multipoles
are denoted by $E_{2-}$, $M_{2-}$ and $S_{2-}$, respectively, and the
transverse multipoles are related to Walker's \cite{walk}
electroproduction helicity amplitudes by,
\begin{eqnarray}
A_{2-} &=& (3M_{2-} -E_{2-})/2 \; , \nonumber \\
B_{2-} &=& M_{2-} + E_{2-} \; .
\end{eqnarray}
In terms of the electroproduction transverse helicity amplitudes, the
transverse helicity amplitudes in the $\gamma ND_{13}$ transition are
given by,
\begin{eqnarray}
A_{1/2}^d &=& C {\rm Im}A_{2-} \; , \nonumber \\
A_{3/2}^d &=& -{ \sqrt{3} \over 2} C {\rm Im} B_{2-} \; ,
\end{eqnarray}
where,
\begin{equation}
C = \left[ { 4 \pi q M_D \Gamma_T^2 \over K_c M \Gamma_{\eta} }
\right] ^{1/2} \; .
\end{equation}
Here, $q$ is the eta three-momentum in the cm frame, $M_D$ is the
D$_{13}$ mass (1520 MeV), $\Gamma_T$ the total D$_{13}$ width,
$\Gamma_{\eta}$ is the D$_{13}$ $\rightarrow$ $\eta$N partial decay
width, and $K_c$ = $(W^2 -M^2)/(2W)$ is the equivalent real photon
energy in the cm frame.
All quantities are to be evaluated at $W$ = $M_D$. For the Coulomb
(or scalar) transition, we have
\begin{equation}
S_{1/2}^d = \sqrt{ { 2Q^2 \over k^2 }} C {\rm Im}S_{2-} \; ,
\end{equation}
again evaluated at $W$ = $M_D$. Here, $k$ is the three-momentum of the
virtual photon in the cm frame.

The parameters of the model are $g_{\eta NN}$, two combinations of
the vector meson couplings, the helicity amplitudes $A_{1/2}^s$ and
$S_{1/2}^s$, the helicity amplitudes $A_{1/2}^d$, $A_{3/2}^d$,
and $S_{1/2}^d$, and the four off-shell parameters, for a
total of 12 parameters. The background is predominantly s- and p-wave,
and controlled by $g_{\eta NN}$, the vector meson couplings and
the off-shell parameters. As there are tremendous correlations amongst
these parameters, we hold $g_{\eta NN}$ and the vector meson couplings
fixed at the values given above and allow the off-shell parameters
to vary in the fits. As this is done at each Q$^2$, the fitted off-shell
parameters can largely compensate for any bias introduced by our
choices of the vector meson couplings and $g_{\eta NN}$.

As the data have practically no angular dependence and the experiment
was performed only at one $\epsilon$, the polarization of the virtual
photon, no separation of $A_{1/2}^s$ and $S_{1/2}^s$ is
possible. The reason for this is that
in the differential cross section, there is no interference
term between the $E_{0+}$ and $S_{0+}$ multipoles\footnote{The $E_{0+}$
multipole is $\sim$ $A_{1/2}^s$, while the $S_{0+}$ multipole is
$\sim$ $S_{1/2}^s$ \protect\cite{ela}.},
and therefore, some arbitrary linear combination
of these two would lead to an angular-independent differential
cross section. At lower Q$^2$, a transverse-longitudinal
separation was made \cite{lt} and it was
found the the longitudinal cross section is small compared to the
transverse one. We assume that this continues to higher Q$^2$, and
do fits with $S_{1/2}^s$ fixed either at zero or roughly\footnote{Specifically,
since we don't expect our result for $A_{1/2}^s$ to be much different
than found in Ref. \cite{arm}, we fix $S_{1/2}^s$ to be 10\% of
$A_{1/2}^s$ found in that work.}
10\% of $A_{1/2}^s$.

Thus, the parameters to be fitted to the data \cite{arm} are
$A_{1/2}^s$, the three helicity amplitudes for the D$_{13}$, and
the four off-shell parameters, for a total of eight parameters.
The results of various fits are given in Table 1 for both Q$^2$ =
2.4 GeV$^2$ and Q$^2$ = 3.6 GeV$^2$.
We use the CERN routine MINUIT to minimize the
chi-squared and the errors on the parameters are the so-called
MINOS errors, which accounts for correlations amongst the
parameters. In all cases, the fit to the data is excellent. In the first
fit, we have fixed $S_{1/2}^s$ at roughly 10\% of
$A_{1/2}^s$ and allowed $A_{1/2}^s$ and the D$_{13}$ parameters (the three
helicity parameters and the off-shell parameters)
to vary.
The extracted $A_{1/2}^s$ is in excellent agreement with
that found in \cite{arm}. At Q$^2$ = 2.4 GeV$^2$, the extracted
helicity amplitudes of the D$_{13}$ all have extremely large errors
and are consistent
with zero. This is consistent with the statement made in Ref. \cite{arm}
that the data are angular independent at the one-sigma level. At Q$^2$
= 3.6 GeV$^2$, there is a slight signal for the presence of the
D$_{13}$. To examine the role of the D$_{13}$ in the fit, we have
turned off the
D$_{13}$ and refitted $A_{1/2}^s$ keeping $S_{1/2}^s$ fixed at
roughly 10 \% of $A_{1/2}^s$. At both Q$^2$ = 2.4 and 3.6 GeV$^2$, the
chi-squared per degree of freedom, $\chi^2$/dgf, {\it increases}, but not
significantly. $A_{1/2}^s$ shifts slightly upward, but within
error is in agreement with that obtained from the first fit. In a final
fit, we examined the role of $S_{1/2}^s$ by setting it zero and
refitting $A_{1/2}^s$ and the D$_{13}$ parameters. The results
turn out to be quite close to the fit with $S_{1/2}^s$ fixed at
10\% of $A_{1/2}^s$.

To summarize our numerical results obtained from the differential
cross section, we find $A_{1/2}^s$
to be 50 $\pm$ 4 $\times$ 10$^{-3}$ GeV$^{-1/2}$ at 2.4 GeV$^2$ and
35 $\pm$ 2 at 3.6 GeV$^2$, in the same units. The errors here do not
take into account uncertainties in the branching ratio, total width and
mass of the S$_{11}$, which were studied in Ref.~\cite{arm}. Our errors
are dominated by uncertainties in the D$_{13}$ contribution, and thus
should be added to the errors found in in Ref.~\cite{arm}.
Our results for $A_{1/2}^s$ are
in excellent agreement with those found in Ref. \cite{arm}. At first
thought, this may come as no suprise since we have used values for the
total width and $\eta$N branching ratio close to their preferred values.
However,
it should be emphasized that our method of analysis is quite different
than the one used in \cite{arm}. In that work, the cross section was
written as an incoherent sum of a resonance contribution and a background
contribution, and it was found that the background contribution was less
than 1\% of the resonant contribution. In our work, we have a coherent
sum of the resonance and background, and find that the background
contributes at the 10\% level or more.
In addition, the role of the D$_{13}$ was
ignored in Ref.~\cite{arm}.

At a $W$ of 1.54 GeV, our results are graphically\footnote{The data shown
in Figs. 1 and 2 represent only a small fraction of the data 
\protect\cite{arm} used in our fit.}
depicted in Fig. 1
at 2.4 GeV$^2$, and in Fig. 2 at 3.6 GeV$^2$. The solid line is the
result of the first fit, while the dashed-line arises when the D$_{13}$ is
turned off, everything else held fixed. These two lines are distinctly
different and the data do seem to favor the solid line, which contains
the D$_{13}$. However, to gauge the {\it need} for the D$_{13}$,
one should compare
with the results of the second fit, that is, with the D$_{13}$ turned off
and $A_{1/2}^s$ refitted to the data. This is shown by the dotted lines
in Figs. 1 and 2. Looking at the graphs, it is difficult to tell if
the data favor the solid or dotted line. At angles where most of the data
exist, the dotted line is roughly the average of the solid line. However,
at Q$^2$ = 2.4 GeV$^2$ and $\phi$ = 90$^{\circ}$, the solid and dotted
lines are quite different. Therefore, data at this angle would help pin
down the D$_{13}$ parameters.

There are two main reasons that the current data do not tightly
constrain the D$_{13}$ helicity amplitudes, in contrast to what happens
at the real photon point \cite{nil}. First, the D$_{13}$ is falling faster
than the S$_{11}$ as a function of Q$^2$, and thus is simply less important
at these Q$^2$ values than at Q$^2$ = 0. Second, the key to pinning
down the D$_{13}$ helicity amplitudes at the real photon point is the
new polarization observables in conjuction with the differential
cross section data. At present, similar polarization observables do not
exist for electroproduction. However, an experiment has recently been
completed \cite{stoler} at the J-Lab, which should have significant
bearing on the extraction of the D$_{13}$ helicity amplitudes from the
eta electroproduction data.

In the J-Lab experiment \cite{stoler}, the data of which are now in the
preliminary stages of analysis,
both the beam and target were longitudinally polarized, i.e.,
parallel or anti-parallel to the beam direction. The polarization of the
target was periodically flipped, resulting in the measurement of
an asymmetry, which we denote as:
\begin{equation}
A_{et}^+ = { \sigma (h=1,p=1) -\sigma (h=1,p=-1) \over
\sigma (h=1,p=1) +\sigma (h=1,p=-1) } \;  ,
\end{equation}
where $h$ is the helicity of the incoming electron and $p$ is the
polarization of the target with $\hat{z}$ \begin {it}defined to be
in the direction of the
incident electron beam.\end{it} Following Ref. \cite{bartl}, but correcting
some mistakes originally pointed out by Dmitrasinovic {\it et al.} \cite{don},
the differential cross section for the case of polarized beam and target
can be written as
\begin{equation}
\sigma (\theta , \phi) = \sigma_0 + \sigma_e + \sigma_t +
\sigma_{et} \; .
\end{equation}
The expression for the
unpolarized cross section, $\sigma_0$, has a standard form and
will not be given here. For the other contributions,
we work with
the transverse helicity amplitudes, $h_{\pm}^i$, which are related to
those of Walker \cite{walk} by
\begin{eqnarray}
h_{\pm}^N &=& (H_4 \pm H_1 ) / \sqrt{2} \; , \\
h_{\pm}^F &=& (H_3 \mp H_1 ) / \sqrt{2} \; .
\end{eqnarray}
Walker's $H_i$ are in turn related to the CGLN \cite{cgln} ${\cal F}$'s by
\begin{eqnarray}
H_{1}(\theta ) &=&-\frac{1}{\sqrt{2}} \sin\theta \cos(\theta /2)
\left[ {\cal F}_{3}+{\cal F}_{4} \right] \\
H_{2}(\theta ) &=& \sqrt{2} \cos(\theta /2)
\left[ {\cal F}_{2}-{\cal F}_{1} +\frac{1}{2}(1-\cos\theta )
( {\cal F}_{3}-{\cal F}_{4} ) \right] \\
H_{3}(\theta )&=& \frac{1}{\sqrt{2}} \sin\theta \sin(\theta /2)
\left[ {\cal F}_{3}-{\cal F}_{4} \right] \\
H_{4}(\theta )&=& \sqrt{2}\sin(\theta /2) \left[ {\cal F}_{2}
+{\cal F}_{1}+\frac{1}{2}(1+\cos\theta )({\cal F}_{3}+
{\cal F}_{4}) \right] .
\end{eqnarray}

The longitudinal helicity amplitudes, $h_0^i$, are related to the
CGLN \cite{don2} ${\cal F}$'s by
\begin{eqnarray}
h_0^N &=& - \sqrt{ { -K^2 \over k^2 } } ( {\cal F}_7 + {\cal F}_8 )
\cos (\theta /2) \; , \\
h_0^F &=& \sqrt{ { -K^2 \over k^2 } } ( {\cal F}_7 - {\cal F}_8 )
\sin (\theta /2) \; .
\end{eqnarray}
Comparing with Ref. \cite{tiator}, our ${\cal F}_{1,2,3,4}$ are the
same as their's, and our ${\cal F}_{7,8}$ are related to their
${\cal F}_{5,6}$ by,
\begin{eqnarray}
k_0 {\cal F}_7 &=& k {\cal F}_6 \; , \\
k_0 {\cal F}_8 &=& k {\cal F}_5 \; ,
\end{eqnarray}
where $k_0$ is the energy of the virtual photon in the cm frame.

For a polarized beam, $\sigma_e$ enters the cross section, and
is given by,
\begin{eqnarray}
\sigma_e &=& h {q \over K_c} \sqrt{2\epsilon (1-\epsilon)} \sin (\phi )
{\rm Im} (h_0^N h_-^{*N}+ h_0^F h_-^{*F} ) \; , \nonumber \\
 &=& h f_e
\end{eqnarray}
For a polarized target, $\sigma_t$ enters and is given by
\begin{eqnarray}\label{sigt}
\sigma_t &=& {q \over K_c } \left[ P_x (\sqrt{2\epsilon (1+\epsilon)}
\sin (\phi ) {\rm Im}X_1 + \epsilon \sin (2\phi) {\rm Im}X_2 )
\right. \nonumber \\
 &-& P_y ( {\rm Im}Y_1 + \epsilon \cos (2\phi) {\rm Im}Y_2 +
 2\epsilon {\rm Im}Y_3 + \sqrt{2\epsilon (1+\epsilon)} \cos (\phi )
{\rm Im}Y_4 ) \nonumber \\
 &-& P_z \left. ( \epsilon \sin (2\phi ) {\rm Im}Z_2 +
\sqrt{2\epsilon (1+\epsilon)} \sin (\phi ) {\rm Im}Z_1 ) \right] \; ,
\nonumber \\
 &=& P_x f_{tx} -P_y f_{ty} -P_z f_{tz} \; ,
\end{eqnarray}
where
\begin{eqnarray}
X_1 &=& h_0^F h_+^{*N}+ h_0^N h_+^{*F}  \qquad
X_2 = h_-^F h_+^{*N}+ h_-^N h_+^{*F}  \nonumber \\
Y_1 &=& h_+^N h_+^{*F}+ h_-^N h_-^{*F}  \qquad
Y_2 = h_-^N h_-^{*F}- h_+^N h_+^{*F}  \nonumber \\
Y_3 &=& h_0^N h_0^{*F}  \qquad \qquad \qquad \hspace{2pt}
Y_4 = h_0^N h_-^{*F}- h_0^F h_-^{*N}  \nonumber \\
Z_1 &=& h_0^N h_+^{*N}- h_0^F h_+^{*F}  \qquad
Z_2 = h_-^N h_+^{*N}- h_-^F h_+^{*F}  \; .
\end{eqnarray}

If both beam and target are polarized, $\sigma_{et}$ also enters:
\begin{eqnarray}\label{siget}
\sigma_{et} &=& h{q \over K_c } \left[ P_x (\sqrt{2\epsilon (1-\epsilon)}
\cos (\phi ) {\rm Re}X_1 + \sqrt{1-\epsilon^2} {\rm Re}X_2 ) \right.
\nonumber \\
 &+& P_y \sqrt{2\epsilon (1-\epsilon)} \sin (\phi )
{\rm Re}Y_4 ) \nonumber \\
 &-& P_z \left. ( \sqrt{1-\epsilon^2} {\rm Re}Z_2 +
\sqrt{2\epsilon (1-\epsilon)} \cos (\phi ) {\rm Re}Z_1 ) \right] \; ,
\nonumber \\
 &=& h(P_x f_{etx} +P_y f_{ety} -P_z f_{etz} ) \; .
\end{eqnarray}

The $P_i$ in (\ref{sigt},\ref{siget}) are defined in a frame
in which $\hat{z}$ \begin{it}is in the direction of the virtual
photon,\end{it} $\hat{y}$ is in the direction of $\vec{k}\times\vec{q}$,
and $\hat{x}$ is in the direction of $\vec{y}\times\vec{z}$. Here, $\vec{k}$
is the three-momentum of the virtual photon and $\vec{q}$ is the
three-momentum of the eta, both in the cm frame.
If the
target is 100\% polarized in the beam direction, then
we find that the relevant
$P_i$ to be used in (\ref{sigt},\ref{siget}) are
\begin{equation}\label{pol}
P_z = \cos \beta \qquad P_x = \sin \beta \cos \phi \qquad
P_y = -\sin \beta \sin \phi \; ,
\end{equation}
where $\beta$ is the angle between the incident beam and the direction
of the virtual photon;
\begin{equation}
\cos \beta = { \nu +Q^2/(2E) \over \sqrt{\nu^2 +Q^2} } \; ,
\end{equation}
with $E$ the lab beam energy and $\nu$ the energy of the virtual photon in
the lab frame.
For the J-Lab experiment, we find $A_{et}^+$ becomes
\begin{equation}
A_{et}^+ = { P_x (f_{tx}+f_{etx}) +P_y (f_{ety}-f_{ty}) -P_z (f_{tz}+
f_{etz} ) \over \sigma_0 + f_e } \; ,
\end{equation}
with the $P_i$ given in (\ref{pol}).

For the case of a pure $E_{0+}$ amplitude, i.e., total S$_{11}$ dominance,
one obtains the simple result
\begin{equation}
A_{et}^+ = -\cos \beta \sqrt{1-\epsilon^2} \; ,
\end{equation}
which is obviously independent of $\theta$ and $\phi$. For the
differential cross-section, it is difficult to isolate the D$_{13}$ since,
with the current statistics, a rescaling of the $E_{0+}$ can mimic
(within the error bars) the effect of the D$_{13}$. On the other hand,
for $A_{et}^+$ a scaling of $E_{0+}$ essentially leaves this observable
unchanged. Therefore, this observable should be able to distinguish
between the fits with and without the D$_{13}$. This is verified, as is shown
in Fig. 3. The solid line is the fit with the D$_{13}$, while the dashed
line is this fit with the D$_{13}$ turned off. The dotted line, which lies
practically on top of the dashed line, is the best fit without the D$_{13}$.
Thus, we see that this observable is very sensitive to the D$_{13}$ and should
provide powerful constraints on its helicity amplitudes.

In summary, we have investigated the role of the D$_{13}$ in various
eta electroproduction observables, which are readibly measurable,
for example, in Hall B at J-Lab. The signal for the D$_{13}$ contribution
is very weak in the present differential cross section data. A full
4$\pi$ coverage of the differential cross section should show increased
sensitivity to the D$_{13}$. Even stronger constraints on the D$_{13}$
should be provided by the asymmetry $A_{et}^+$, discussed above, which
has been recently measured at J-Lab. Regarding the S$_{11}$, we confirm
the results of Ref.~\cite{arm}, but the errors in that work should be
increased slightly due to uncertainties in the D$_{13}$ sector. Finally,
a tranverse-longitudinal separation would be useful as the Q$^2$ dependences
of the longitudinal amplitudes are also of interest as tests of QCD-inspired
models, and of QCD itself.

We are grateful to C. Armstrong and P. Stoler for
providing us their
results on eta electroproduction prior to publication. We also thank
P. Stoler for a critical reading of this manuscript.
Our research is
supported by the U. S. Department of Energy.

\begin{table}[h]
\caption{The helicity amplitudes extracted from the data
\protect\cite{arm} using our
effective Lagrangain approach. At each Q$^2$, the first row gives the
results when $A_{1/2}^s$ and the D$_{13}$ parameters are allowed to
vary with S$_{1/2}^s$ fixed at roughly 10\% of $A_{1/2}^s$. The second
row shows the best fit with the D$_{13}$ turned off. The last row shows
the best fit under the assumption S$_{1/2}^s$ = 0.}
\vspace{0.1in}
\begin{tabular}{|c|c|c|c|c|c|c|}
Q$^{2}$ & \multicolumn{2}{c|}{$N^{*}(1535)$} &
\multicolumn{3}{c|}{$N^{*}(1520)$} &
\multicolumn{1}{c|} {$\chi^{2}$/dgf} \\
\cline{2-6}
GeV$^{2}$&$A_{1/2}^s$  & $S_{1/2}^s$  & $A_{1/2}^d$  &
$A_{3/2}^d$ & $S_{1/2}^d$  & 
\\ 
\hline
 & 49 $\pm$ 2 & 5.0 & -2 $\pm$ 66 & 96 $\pm$ 99 & 24 $\pm$ 91 & 0.80 \\
2.4& 51.5 $\pm$ 0.3 &5.0&0.0&0.0&0.0&0.88\\
 & 49.5 $\pm$ 3.5 & 0.0 & 5 $\pm$ 76 & 84 $\pm$ 113 & 19 $\pm$ 102 & 0.80 \\
\hline
 & 34 $\pm$ 1 & 3.5 & 18 $\pm$ 8 & 5 $\pm$ 11 &-13 $\pm$ 9 & 0.79\\
 3.6 & 36.7 $\pm$ 0.2 &3.5&0.0&0.0&0.0&0.87\\
 & 35 $\pm$ 1 & 0.0 & 18 $\pm$ 6 & 0 $\pm$ 5 & -14 $\pm$ 8 & 0.79\\
\end{tabular}
\end{table}

 
\begin{figure}[h]
\caption{Comparison of our fits to the J-Lab data \protect\cite{arm}
at $W$ = 1.54 GeV and Q$^2$ = 2.4 GeV$^2$. Solid line is the fit where the
D$_{13}$ parameters vary and it is assumed (S$_{1/2}^s$)/(A$_{1/2}^s$)
$\approx$ 10\%. The
dashed-line is obtained from this fit when the D$_{13}$ is turned off,
everything else held fixed. The dotted line is the best fit without
the D$_{13}$.}
\end{figure}

\begin{figure}[h]
\caption{Comparison of our fits to the J-Lab data \protect\cite{arm}
at $W$ = 1.54 GeV and Q$^2$ = 3.6 GeV$^2$. Curves as in Fig. 1.}
\end{figure}


\begin{figure}[h]
\caption{Predictions for $A_{et}^+$, defined in the text,
at $W$ = 1.54 GeV based on fits to the
differential cross section. Curves as in Fig. 1.}
\end{figure}

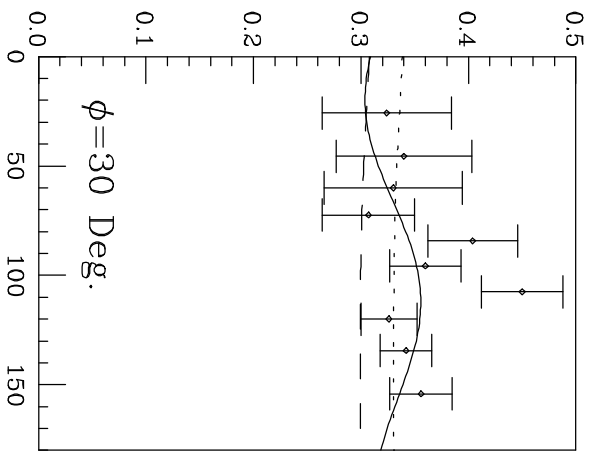
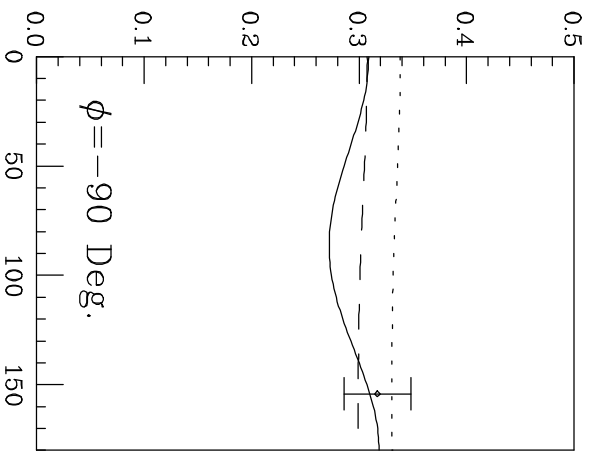
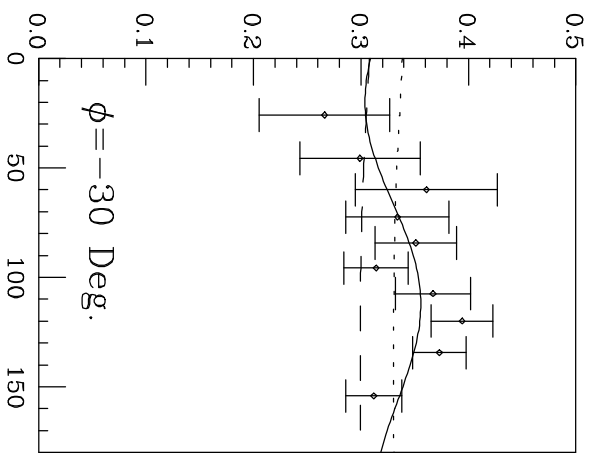
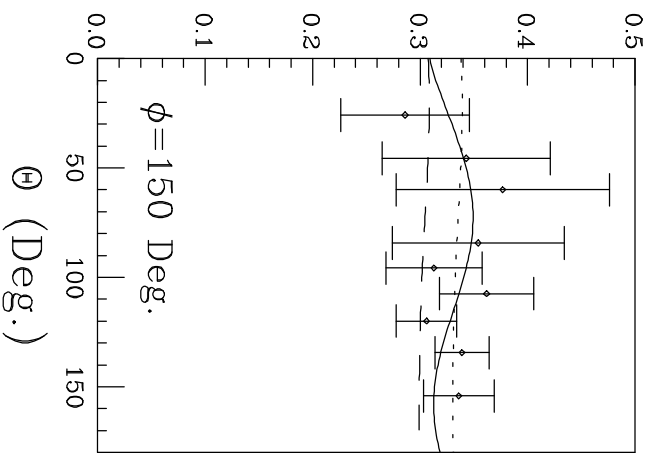
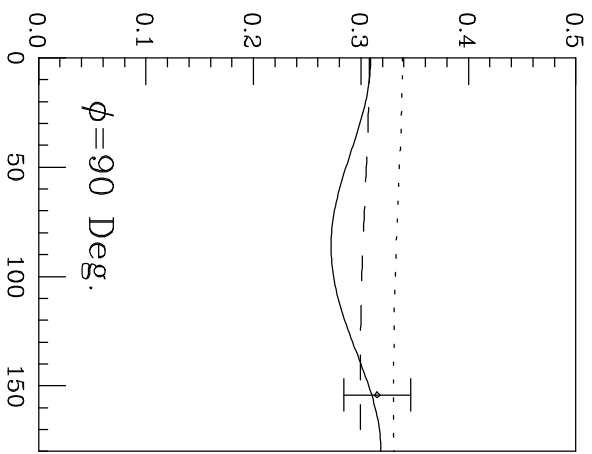
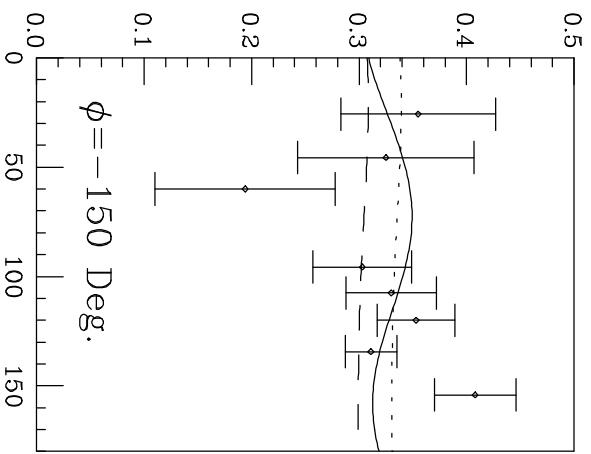
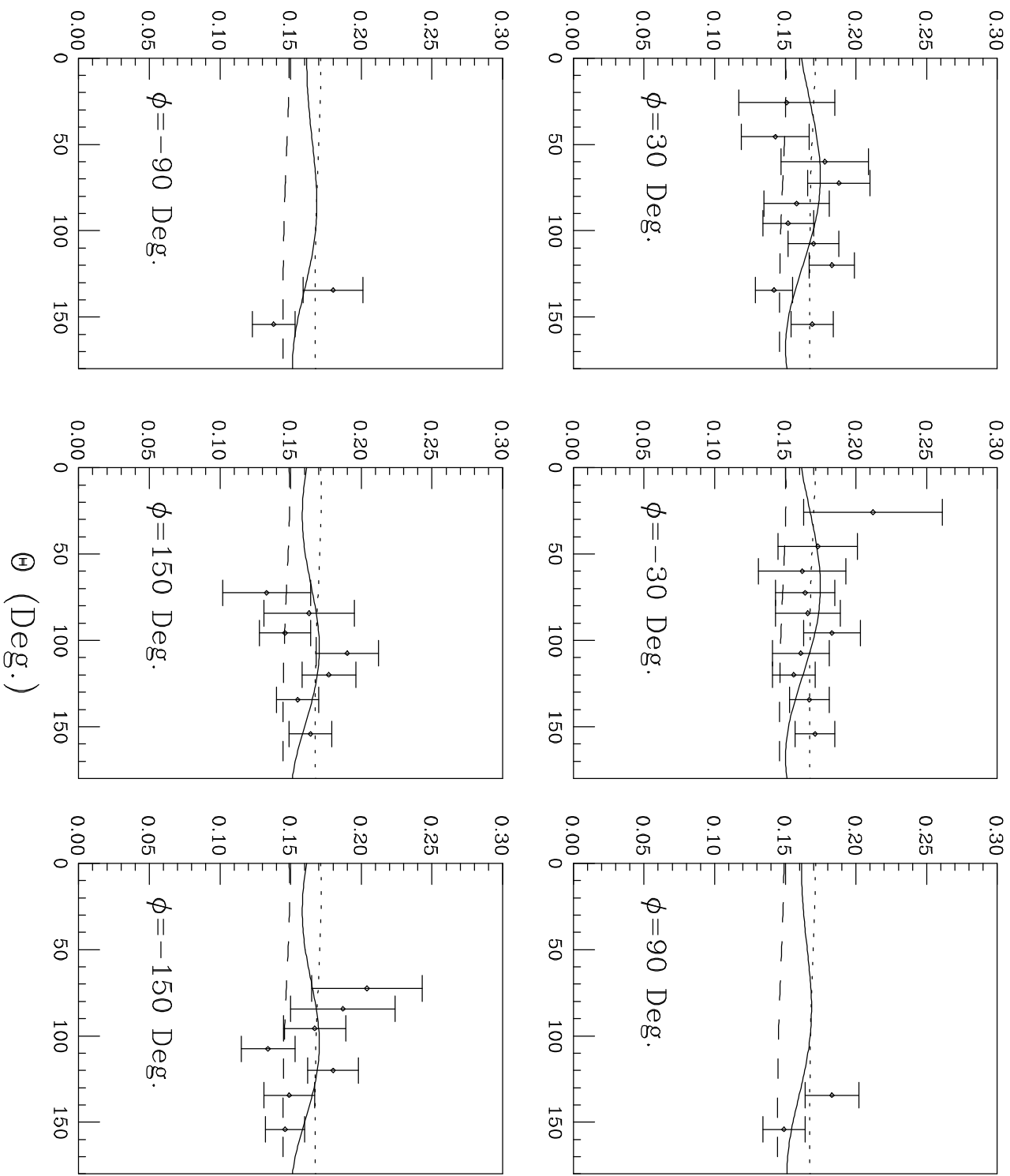
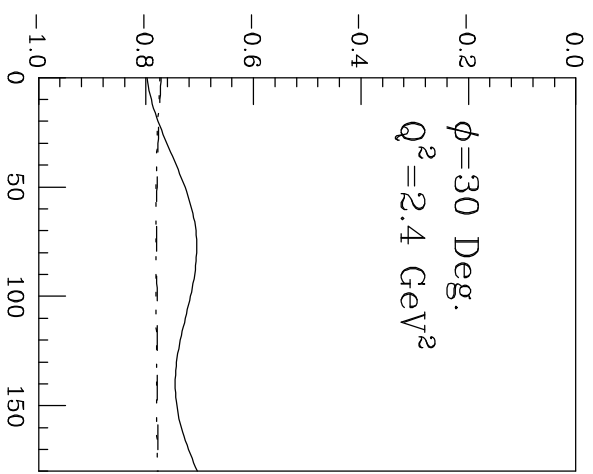
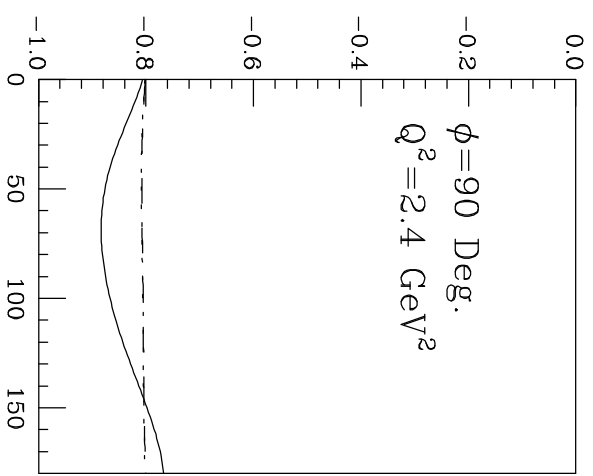
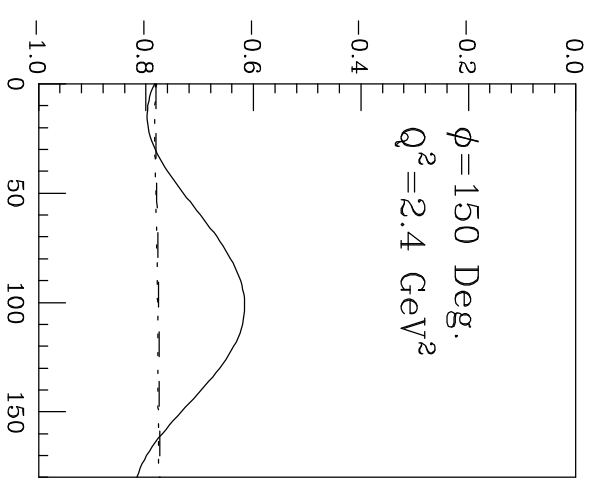
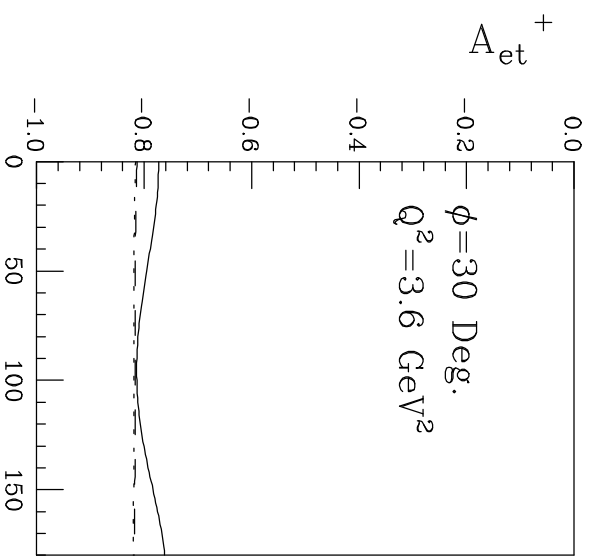
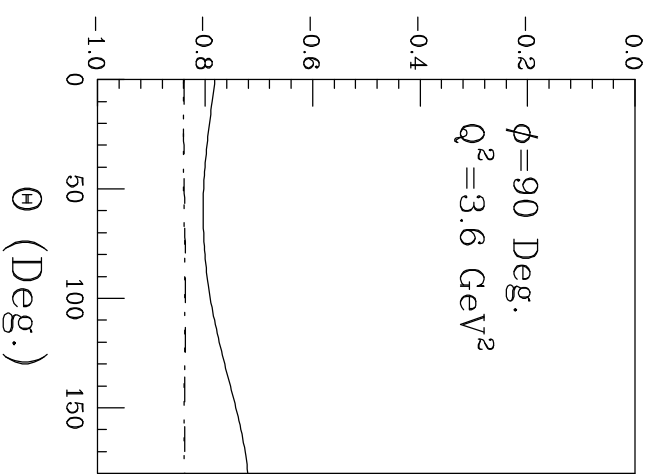
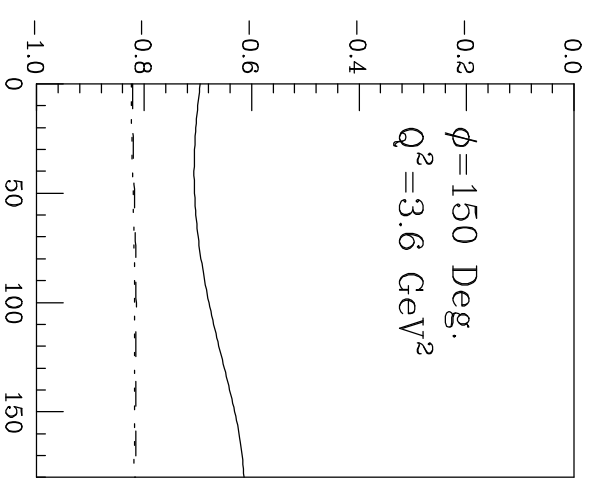
\end{document}